\documentclass[onecolumn,11pt]{article}
\usepackage{jheppub}

\usepackage[font={small}]{caption}
\usepackage{subfigure}
\usepackage{graphicx}
\usepackage{dcolumn}
\usepackage{amsfonts}
\usepackage{amssymb}
\usepackage{bbold}

\makeatletter
\renewcommand{\@seccntformat}[1]{\csname the#1\endcsname.\,\,}
\makeatother
\allowdisplaybreaks

\usepackage{xcolor}
\definecolor{vub}{RGB}{0,52,154}
\usepackage{xargs}
\usepackage[colorinlistoftodos,prependcaption,textsize=tiny]{todonotes}
\newcommandx{\unsure}[2][1=]{\todo[linecolor=vub,backgroundcolor=vub!25,bordercolor=vub,#1]{#2}}

\usepackage{float}
\usepackage[normalem]{ulem}

\usepackage{mathrsfs} 

\usepackage{multicol}
\usepackage{cancel}
\usepackage{mathtools}
\usepackage{hyperref}

\usepackage{makecell}
\usepackage{braket}

\usepackage{subfigure}
\usepackage{xcolor}
\usepackage{soul}

\def\({\left(} \def\){\right)}
\def\[{\left[} \def\]{\right]}

\newcommand{\tr}{\text{Tr}}

\newcommand{\bea}{\begin{eqnarray}}
\newcommand{\eea}{\end{eqnarray}}
\def\p{\partial}
\usepackage{changepage}

\newcommand{\spa}{\ , \ \ }

\newcommand{\beq}{\begin{equation}}
\newcommand{\eeq}{\end{equation}}

\newcommand{\dd}{\mathrm{d}}

\def\tr           {\mathop{\rm Tr}}

\renewcommand{\eqref}[1]{(\ref{#1})}

\newcommand{\CO}{\mathcal{O}}

\title{Gravitational solitons and non-relativistic string theory}
\author{Troels Harmark$^{1, \, 2}$, Johannes Lahnsteiner$^{2}$, Niels A. Obers$^{1, \, 2}$ }

\affiliation{$^1$Niels Bohr International Academy, Niels Bohr Institute, University of Copenhagen, \\
Blegdamsvej 17, DK-2100 Copenhagen Ø, Denmark}
\affiliation{$^2$Nordita, KTH Royal Institute of Technology and Stockholm University, \\
Hannes Alfvéns väg 12, SE-106 91 Stockholm, Sweden}

\emailAdd{harmark@nbi.ku.dk}
\emailAdd{johannes.lahnsteiner@su.se}
\emailAdd{obers@nbi.ku.dk}

\abstract{We explore the non-relativistic string theory (NRST) limit of type II string theory and its action on gravitational solitons.
As a start, we exhibit in detail that the NRST limit is T-dual to a discrete lightcone limit and can be viewed as a near-BPS limit. This also clarifies the nature of multi-string states of NRST and its connection to matrix string theory.
We consider the NRST limit of the fundamental string soliton, confirming the recent finding that it corresponds to a relativistic near-horizon background, which we argue is the manifestation of 
a  strong coupling phase of the NRST worldsheet theory.
Furthermore, we consider the NRST limit of a class of D-branes as well as the NS5-brane. This reveals that they become gravitational solitons in NRST, as they are sourced torsional string Newton-Cartan (TSNC) geometries.
Finally, for the NRST D-brane solitons we show that a further decoupling limit leads to new holographic correspondences between multicritical matrix theories and NRST in curved TSNC backgrounds.}

\begin{document}

\maketitle

\setcounter{tocdepth}{1}

\section{Introduction}

Non-relativistic string theory has evolved in recent years into a mature field, being a consistent corner of string theory  opening  new avenues for exploring holography and non-perturbative phenomena. Originally introduced almost 25 years ago \cite{Klebanov:2000pp,Gomis:2000bd,Danielsson:2000gi,Danielsson:2000mu}, 
the framework emerges from a non-relativistic limit of relativistic string theory, analogous to the non-relativistic limit of relativistic particles. Importantly, non-relativistic string theory (NRST) is a consistent sector within relativistic string theory, as has been  demonstrated through 
string scattering amplitudes \cite{Gomis:2000bd,Yan:2021hte}, 
sigma-model formulations 
\cite{Harmark:2017rpg,Kluson:2018egd,Bergshoeff:2018yvt,Harmark:2018cdl,Harmark:2019upf,Bergshoeff:2019pij,Bidussi:2021ujm}, 
 beta-function analyses \cite{Gomis:2019zyu, Gallegos:2019icg, Yan:2019xsf, Bergshoeff:2021bmc, Yan:2021lbe} and supergravity constructions \cite{Bergshoeff:2021tfn, Bergshoeff:2023ogz,Bergshoeff:2024nin}. 

Despite  the name,  non-relativistic string theory in its original formulation does not exhibit non-relativistic symmetries on the worldsheet. 
It is rather the target space spectrum, the S-matrix as well as the curved background geometry that carry non-relativistic symmetries. In particular the appropriate target space geometry is a type of non-Lorentzian geometry called (torsional) string Newton-Cartan geometry \cite{Andringa:2012uz,Harmark:2017rpg,Bergshoeff:2018yvt,Bidussi:2021ujm},  
 which is a string generalization of  Newton-Cartan geometry relevant for non-relativistic particles (see \cite{Oling:2022fft,Bergshoeff:2022eog,Hartong:2022lsy,Demulder:2023bux} for various reviews). 
However, as has become clear as well over the last years, when implementing further limits, e.g. along the longitudinal direction of the string 
\cite{Harmark:2017rpg,Harmark:2018cdl, Harmark:2019upf, Harmark:2020vll,Roychowdhury:2020yun,Roychowdhury:2021wte,Bidussi:2021ujm, Bidussi:2023rfs,Blair:2023noj,Gomis:2023eav}, non-Lorentzian worldsheet structures can emerge as well. 

The non-relativistic limit can be taken in bosonic string theory, where it relies crucially on the property that the tension and charge of a fundamental string are equal \cite{Bidussi:2021ujm} along with a critical $B$-field. 
In the context of superstring theory, the non-relativistic limit is even more natural as it corresponds to the near-BPS limit, also known as the BPS decoupling limit, of the fundamental string
\cite{Blair:2023noj,Gomis:2023eav}. 
This feature
not only solidifies its foundational role but also links it to the dynamics of supersymmetric sectors of string theory.

Indeed, recent developments have highlighted the role of NRST within a broader web%
\footnote{The classification of  BPS decoupling limits into duality orbits linked to one or multiple DLCQs was also 
considered in \cite{bpslimits} using U-duality invariant mass formulae  (for a review see \cite{Obers:1998fb}).}
of BPS decoupling limits related by dualities in string theory \cite{Blair:2023noj,Gomis:2023eav,Blair:2024aqz}. This shows that non-relativistic strings are intimately linked to matrix theories, which correspond to analogous BPS decoupling limits  associated with D-branes.%
\footnote{See \cite{Taylor:2001vb} for a review on matrix theory. Recently, new bootstrap techniques
\cite{Han:2020bkb, Lin:2023owt,Maldacena:2023acv, Biggs:2023sqw,Komatsu:2024vnb, Cho:2024kxn, Lin:2024vvg}
and amplitude studies \cite{Miller:2022fvc, Tropper:2023fjr,Herderschee:2023pza,Herderschee:2023bnc}  have revived this subject, which is thus further informed by the new developments in NRST.} 
Beyond matrix theory, the duality web furthermore includes links to holography, spin matrix theory \cite{Harmark:2014mpa} as well as tensionless \cite{Lindstrom:1990qb, Isberg:1993av,Bagchi:2013bga} and Carrollian strings \cite{Cardona:2016ytk,Blair:2023noj,Bagchi:2023cfp,Harksen:2024bnh,Bagchi:2024rje}. 
 These relationships provide a unified perspective on how different sectors of string theory emerge from consistent limiting procedures, reinforcing the centrality of non-relativistic strings in the broader framework of string theory and in the study of non-Lorentzian holography. 

The purpose of this paper is to explore gravitational solitons of type II string theory in the NRST limit,%
\footnote{See also \cite{Bergshoeff:2022pzk} which obtains various NRST brane solutions in the NS sector using longitudinal T-duality as a solution generating technique.}  
their connection to perturbative and non-perturbative NRST, and their role in holographic dualities.
In relativistic string theory gravitational solitons are central to understanding the intricate interplay between geometry and dynamics as exemplified beautifully in the AdS/CFT correspondence.  A natural question arises: what becomes of these solitonic solutions when the NRST limit  is taken? 
Exploring this, novel gravitational solitons as well as holographic correspondences within NRST are revealed. Moreover, it leads to important insights into to the  NRST worldsheet theory.

Since the NRST limit is central to this paper, we begin by 
reviewing and revisiting its formulation, focussing in particular on its presentation as a 
discrete light cone quantization (DLCQ) followed by a T-duality \cite{Harmark:2017rpg,Kluson:2018egd,Bergshoeff:2018yvt}. To make this procedure well-defined and operational, we implement the DLCQ  \cite{Susskind:1997cw} 
 following \cite{Seiberg:1997ad,Sen:1997we}, as the compactification on a boosted spatial circle, which becomes lightlike in the NRST limit. 
In this procedure, the winding mode of the NRST limit is dual to the momentum mode in the DLCQ on the T-dual side.  
 This approach ensures that the T-duality remains well-defined before taking the limit, enabling a rigorous exploration of the behavior of gravitational solitons under the non-relativistic limit.

This way of approaching NRST also sheds light on the nature of multi-string states in non-relativistic string theory (NRST) and their relationship with matrix string theory \cite{Motl:1997th,Dijkgraaf:1997vv}. The latter arises from dualizing BFSS matrix quantum mechanics  \cite{Banks:1996vh,Susskind:1997cw} on a spatial circle and 
can thus be considered to be a second quantization of non-relativistic strings.  

The first gravitational soliton we consider is the fundamental string soliton which can be seen as the backreacted winding mode. 
Taking the NRST limit of this soliton we confirm the recent finding \cite{Avila:2023aey} that it corresponds to a relativistic near-horizon background, due to the infinite backreaction in the limit. Building on the original work \cite{Maldacena:1997re,Itzhaki:1998dd}, we argue that the near-horizon background is the manifestation of a strong coupling phase of the NRST worldsheet theory.  The fact that the NRST limit becomes a near-horizon limit parallels the results of \cite{Blair:2024aqz} which considered the analogous D-brane BPS decoupling limits leading to matrix theories. 

Turning to other gravitational solitons of type II string theory, we find examples of solitons
for which the NRST limit does not lead to infinite backreaction. 
Hence, this provides genuine gravitational solitons in NRST described by appropriate sourced (torsional) Newton-Cartan (TSNC) geometries. As we detail in this paper, this is the case for transverse D-branes and NS5-branes, where transverse means in relation to the longitudinal spatial direction involved in the NRST limit.%
\footnote{Note our transverse D-brane soliton is smeared on the T-dual circle. See \cite{Guijosa:2023qym,Avila:2023aey} for a study of the non-smeared case.}
We expect that the transverse D-brane soliton corresponds to the
backreacted transverse D-brane state of the open string sector of NRST 
\cite{Gomis:2020fui,Gomis:2020izd,Kluson:2020kyp,Ebert:2021mfu,
Guijosa:2023qym,Hartong:2024ydv}.
The construction of the NRST D-brane is treated in detail, and from this we infer a general procedure to generate further NRST geometries given relativistic seed solutions that satisfy appropriate conditions. This is illustrated by applying it to the NS5-brane case, for which the resulting NRST NS5-brane coincides with the soliton solution obtained in Ref.~\cite{Bergshoeff:2022pzk} using another method.  

These new solitons of NRST are 
especially intriguing as they allow taking further near-horizon limits. 
In particular, we show that for NRST D-brane solitons the near-horizon limit corresponds to a further decoupling limit on the boundary. This generates  novel   holographic correspondences between multicritical matrix brane theories \cite{Blair:2023noj,Gomis:2023eav,Blair:2024aqz} and NRST in a particular curved TSNC backgrounds. 
These boundary theories, involving multicritical BPS decoupling limits, are referred to as  MM$p$T.

To show that this is the relevant boundary theory, we invoke the results of \cite{Blair:2024aqz}, in which it was shown that for (relativistic) D$p$-branes the near-horizon limit corresponds to the matrix $p$-brane (M$p$T) limit on the boundary, coupling to $p$-brane Newton-Cartan geometry. In our case, we need to apply this to a D$(p+1)$-brane and combine this with the additional NRST limit. But this is precisely  MM$p$T, being related to M$(p+1)$T via a further DLCQ and subsequent T-duality. Worldvolume actions for strings and D-branes in MM$p$T have been given in \cite{Gomis:2023eav} whereas the non-Lorentzian target space geometry is presented in \cite{Blair:2023noj}.

The holographic correspondences of this paper thus parallel those obtained in \cite{Blair:2024aqz} and fall into the paradigm that string theory holographic dualities can be viewed as DLCQ${}^m$/DLCQ${}^n$ correspondences with $m<n$. Here,  AdS/CFT \cite{Maldacena:1997re,Itzhaki:1998dd} is of the ${\rm DLCQ}^0/{\rm DLCQ}^1$ type, which thus also includes the duality between strongly coupled NRST worldsheet theory and the near-horizon F-string soliton discussed in this paper. 
Moreover, the duality involving the near-horizon NRST D-brane and hence an extra DLCQ on the boundary, comprises yet another compelling example of the recently found  ${\rm  DLCQ}^1/{\rm DLCQ}^2$ type of holographic dualities. Examples of this type associated to D-brane bound states
are given in \cite{Lambert:2024yjk,Blair:2024aqz,Lambert:2024ncn}
and similar ones using membranes in \cite{Lambert:2024uue}.%
\footnote{For another proposal of holography involving non-relativistic strings see \cite{Fontanella:2024rvn, Fontanella:2024kyl} as well as the earlier non-relativistic string limit of AdS${}_5 \times S^5$  \cite{Gomis:2005pg}. In those works the critical $B$ field is added by hand, and not obviously part of the BPS structure.} 
More generally, this class follows from consecutively applying two  BPS decoupling limits, or equivalently a multi-critical limit, enabling the generation of additional examples of holography, including novel frameworks with non-Lorentzian bulk geometries. 

The results of this paper thus deepen the understanding of the duality web surrounding NRST and its associated gravitational solitons, while also uncovering new connections to holography and novel boundary theories.

\section{What is non-relativistic string theory?}
\label{sec_NRST}

In this section we provide two illuminating perspectives on non-relativistic string theory (NRST).
In Section \ref{sec:near_BPS}
we explain how the
NRST limit can be seen as a near-BPS limit (BPS decoupling limit). This is done through a T-duality argument which also relates the limit to the discrete light cone quantization (DLCQ) of a string.
In Section \ref{sec:twisted_sectors} we discuss the relation between NRST and matrix string theory, providing an understanding of second quantization of NRST.

\subsection{A near-BPS limit of string theory}
\label{sec:near_BPS}

In this section we exhibit how the non-relativistic string theory (NRST) limit is associated with the DLCQ of strings through a standard T-duality along a spatial direction. 
This in turn makes it clear in which sense the NRST limit can be seen as a near-BPS limit.
The section uses arguments from the original papers \cite{Klebanov:2000pp,Gomis:2000bd,Danielsson:2000gi,Danielsson:2000mu} as well as the more recent take on NRST \cite{Harmark:2017rpg,Bergshoeff:2018yvt,Harmark:2018cdl,Harmark:2019upf} (see also the review \cite{Oling:2022fft}).

To begin, we consider flat ten-dimensional space-time with a compact direction $\mathbb{R}^{1,8} \times S^1$. The metric is 
\begin{equation}
\label{metric_orig}
ds^2 = -dT^2 + dX^2 + dr^i dr^i \ , 
\end{equation}
where we sum over $i= 1,2,...,8$. We take $X$ to be the compact direction of radius $R_X$. In addition we have all other fields being zero, including the Kalb-Ramond field $B=0$ and the dilaton $\Phi=0$. The string coupling is written as $\bar{g}_s e^\Phi = \bar{g}_s$ and is assumed to be small enough that one can consider a single string.

We consider a single relativistic closed string  in this background with a non-zero momentum along the compact direction $X$
\begin{equation}
p_X = \frac{w}{R_X} \ , 
\end{equation}
with $w$ a positive integer, and a possible winding number $n$ along $X$.
The mass formula for the string is
\begin{equation}
\label{massform_origframe}
    E_T^2- \frac{w^2}{R_X^2} - \delta^{ij} p_i p_j =   n^2 \frac{R_X^2}{(\alpha')^2} + \frac{2}{\alpha'} ( N+ \bar{N}-a) \ , 
\end{equation}
where $E_T = - (\partial_T)^\mu p_\mu = - p_T$ is the energy, $p_i$ are the momenta of the eight transverse directions and $\alpha'$ is the string length squared.
The mass formula  includes 
possible left- and right moving excitations $N$ and $\bar{N}$ of the string, subject to a level matching condition, as well as a normal-ordering constant $a$ depending on the boundary conditions of the string. 

The string state with $E_T = p_X = \tfrac{w}{R_X}$ is a $1/2$-BPS state. The goal in the following is to take a near-BPS decoupling limit that zooms in on the modes which corresponds to small excitations above this state.
To this end, we introduce the new coordinates $(t,u)$ as \cite{Bilal:1998vq}
\begin{equation}
\label{coords_tu}
T = c\, t \spa  X = c\, t + \frac{1}{c} u \ , 
\end{equation}
where $c$ is an arbitrary positive dimensionless parameter. For large $c$ and finite $(t,u)$ one approaches the edge of the light cone in $(T,X)$ coordinates.
The metric takes the form
\begin{equation}
\label{metric_tu}
ds^2 =  2dt \, du + \frac{1}{c^2} du^2  + dr^i dr^i \ , 
\end{equation}
still with $B=0$ and $\bar{g}_s e^\Phi = \bar{g}_s$.
In these coordinates $u$ is compact with radius $R_u = c \, R_X$. The closed string mass formula becomes
\begin{equation}
\label{massform_DLCQframe}
    \frac{1}{c^2} E_t^2  + 2 E_t \frac{w}{R_u} - \delta^{ij} p_i p_j= \frac{1}{c^2} \left( \frac{nR_u}{\alpha'}\right)^2 + \frac{2}{\alpha'} (N+\tilde{N}-a) \ . 
\end{equation}
Here $E_t = - (\partial_t)^\mu p_\mu = - p_t$ is the energy.
Notice that one has
\begin{equation}
\label{Et_formula}
    E_t = c\,  ( E_T - p_X ) \spa p_u = \frac{1}{c} p_X \ . 
\end{equation}
Taking now the limit \cite{Klebanov:2000pp,Gomis:2000bd,Danielsson:2000gi}
\begin{equation}
\label{limit_DLCQ}
c\rightarrow \infty \ \mbox{with}\  \alpha' , \ R_u \ \mbox{and} \ (t,u,r^i) \ \mbox{fixed} \ , 
\end{equation}
one gets the energy spectrum
\begin{equation}
\label{massform_limit_DLCQ}
    E_t =    \frac{R_u}{2 w} \left[ \delta^{ij} p_i p_j +\frac{2}{\alpha'} (N+\tilde{N}-a) \right] \ . 
\end{equation}
The $c\rightarrow \infty$ limit \eqref{limit_DLCQ} translates to a DLCQ limit, since the limit of the metric \eqref{metric_tu} is
\begin{equation}
\label{metric_tu_limit}
ds^2 =  2dt \, du   + dr^i dr^i \ , 
\end{equation}
with the $u$ isometry becoming null,  and $R_u$ is finite in the limit. 

\begin{figure}[h]
\begin{picture}(390,75)(-20,0)
\put(5,40){\vector(1,0){380}}
\put(378,50){$E_T$}
\put(2,50){0}
\put(5,35){\line(0,1){10}}
\put(123,55){$\tfrac{w}{R_X}$}
\put(132,35){\line(0,1){10}}
\put(134,37){\line(0,1){6}}
\put(136,37){\line(0,1){6}}
\put(138,37){\line(0,1){6}}
\put(140,37){\line(0,1){6}}
\put(142,37){\line(0,1){6}}
\put(144,37){\line(0,1){6}}
\put(146,37){\line(0,1){6}}
\put(148,37){\line(0,1){6}}
\put(150,37){\line(0,1){6}}
\put(152,37){\line(0,1){6}}
\put(154,37){\line(0,1){6}}
\put(156,37){\line(0,1){6}}
\put(158,37){\line(0,1){6}}
\put(160,37){\line(0,1){6}}
\put(162,37){\line(0,1){6}}
\put(77,15){\shortstack{$\underbrace{\phantom{xxxxx}}$ \\ Modes of DLCQ/NRST limit}}
\end{picture}
\caption{Illustration of the DLCQ/NRST limit in the original frame \eqref{metric_orig} where it is a near-BPS limit. The modes of the DLCQ/NRST limit have energies $E_T = \tfrac{w}{R_X} + \CO(\tfrac{1}{c})$.\label{fig:near_BPS} }
\end{figure}
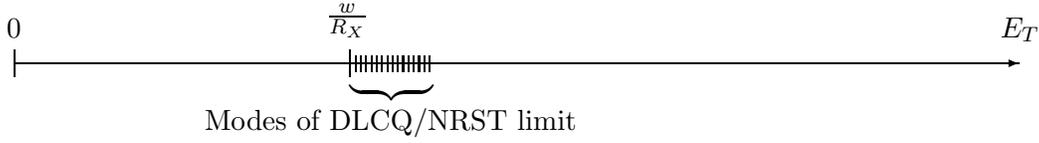

One can see from \eqref{Et_formula} that the $c\rightarrow \infty$ limit corresponds to zooming in on the modes which are close to the 1/2-BPS state $E_T = p_X = w/R_X$ given a choice of $w>0$. Indeed, one considers modes for which $c( E_T - w/R_X)$ is of order one for $c\rightarrow \infty$.
This is illustrated in Figure \ref{fig:near_BPS}. 
Thus, $c\rightarrow \infty$ is a near-BPS limit. 

An alternative point of view on the near-BPS/DLCQ limit \eqref{limit_DLCQ} is found by first performing a
T-duality along the space-like isometry $u$ on the finite-$c$ background given by  \eqref{metric_tu} and $B=\Phi=0$. Employing the standard Buscher rules \cite{Buscher:1987sk,Alvarez:1994dn} we find
\begin{equation}
\label{solution_NRST}
ds^2 =  c^2 (-dt^2 + dv^2 )  + dr^i dr^i \spa B = - c^2 dt\wedge dv \spa  e^{\Phi} =  c \ , 
\end{equation}
where $v$ is the T-dual direction which is compact with radius $R_v$. The proper radius under T-duality is $c \, R_v = \alpha' / ( \tfrac{1}{c} R_u)$ hence we have
\begin{equation}
\label{Ru_Rv}
    R_v = \frac{\alpha'}{R_u} \ . 
\end{equation}
We define
\begin{equation}
\label{bar_gs}
    g_s = \bar{g}_s \frac{\sqrt{\alpha'}}{R_u} \ , 
\end{equation}
such that the full string coupling of the background \eqref{solution_NRST} is
\begin{equation}
\label{NRST_coupling}
    g_s e^\Phi = c\, \bar{g}_s \frac{\sqrt{\alpha'}}{R_u} \ . 
\end{equation}

In the T-dual frame the closed string mass formula is
\begin{equation}
\label{massform_NRSTframe}
\frac{1}{c^2}E_t^2 + 2  E_t \frac{w R_v}{\alpha'} - \frac{1}{c^2} \frac{n^2}{R_v^2}  - \delta^{ij}p_i p_j=      \frac{2}{\alpha'} (N+\tilde{N}-a) \ , 
\end{equation}
with $E_t = - (\partial_t)^\mu p_\mu = - p_t$.

The $c\rightarrow \infty$ limit \eqref{limit_DLCQ} now translates to
\begin{equation}
\label{limit_NRST}
c\rightarrow \infty \ \mbox{with}\  \alpha' , \ R_v \ \mbox{and} \ (t,v,r^i) \ \mbox{fixed} \ , 
\end{equation}
in which the Kalb-Ramond field becomes critical.
This is precisely the non-relativistic string (NRST) limit of \cite{Klebanov:2000pp,Gomis:2000bd,Danielsson:2000gi} in the particular case of a flat-space background.
This limit gives the mass formula
\begin{equation}
\label{massform_limit_NRST}
    E_t =    \frac{\alpha'}{2 w R_v} \left[ \delta^{ij} p_i p_j +\frac{2}{\alpha'} (N+\tilde{N}-a) \right] \ , 
\end{equation}
corresponding to a non-relativistic spectrum with regard to the transverse directions. It is obviously equivalent to \eqref{massform_limit_DLCQ} using \eqref{Ru_Rv}.

Inserting the background \eqref{solution_NRST} in the Nambu-Goto or Polyakov sigma-model for the relativistic string, the $c\rightarrow \infty$ limit can be interpreted as a string that propagates in a 
flat (torsional) string-Newton-Cartan ((T)SNC) background \cite{Harmark:2017rpg,Bergshoeff:2018yvt,Harmark:2018cdl,Harmark:2019upf}
\begin{equation}
\label{eq:flatTSNC}
    \tau_{\mu\nu} dx^\mu dx^\nu = -dt^2+dv^2 \spa E_{\mu\nu} dx^\mu dx^\nu = dr^i dr^i \ , 
\end{equation}
where $\tau_{\mu\nu}$ is the longitudinal metric and $E_{\mu\nu}$ the transverse metric. For this case, there is no torsion  and the so-called torsional string 
Newton-Cartan two-form is zero along with the limiting dilaton field. 
This gives the action of NRST on flat spacetime \cite{Gomis:2000bd}
\begin{equation}
    \label{NRST_worldsheet}
    S_{\rm NRST} = - \frac{1}{4\pi \alpha'} \int d^2 \sigma \, \eta^{\alpha\beta}
    \partial_\alpha X^i \partial_\beta X^i
\end{equation}
written in gauge-fixed form.
In Section \ref{sec:branes_NRST}  more general torsional string-Newton-Cartan (TSNC)  backgrounds are considered which include torsion, as well as a non-trivial TSNC two-form and dilaton field.

We note that the full string coupling $g_s e^\Phi$ in \eqref{NRST_coupling} goes to infinity. However, the surviving modes can be shown to couple to a rescaled dilaton field which is finite in the limit. For this particular background it means the weak coupling requirement reduces to $g_s \ll 1$. Consistent with this, note also that the string is weakly coupled in the T-dual frame upon taking the limit.

The above T-duality argument shows explicitly that the string background \eqref{solution_NRST} for large but finite $c$ is naturally mapped via a space-like T-duality to \eqref{metric_tu} which is close to a null reduction. This provides a finite-$c$ realization of the relation between the (torsional) string-Newton-Cartan and null reduction approaches to NRST \cite{Harmark:2017rpg,Bergshoeff:2018yvt,Harmark:2018cdl,Harmark:2019upf}.%
\footnote{The finite $c$ versions of  T-duality Buscher rules and the limiting procedure were also discussed in \cite{Bergshoeff:2019pij,Ebert:2021mfu}
and related versions have appeared in \cite{Lahnsteiner:2022mag,2023:harmark_unpublished,Blair:2023noj}.}

\subsection{Multi-string states and interactions in non-relativistic string theory}
\label{sec:twisted_sectors}

An important property of non-relativistic string theory is that on-shell physical 
single-string states carry positive  winding along the $v$-direction \cite{Gomis:2000bd,Danielsson:2000gi}. This means in turn that the multi-string states with total winding number $N$ are characterized by the possible partitions of $N$ into the sum of positive integers. Moreover, since the total winding number should be conserved in string scattering processes, this has a direct consequence for string perturbation theory. Namely, one can see the topological expansion 
as arising from an interaction that incorporates the joining and splitting of wound strings such that the total winding number is conserved in the process.  

In fact, for type IIB NRST this physical expectation 
is beautifully realized in the 
matrix string theory proposal~\cite{Motl:1997th, Dijkgraaf:1997vv}, which indeed corresponds to the same near-BPS decoupling limit of the worldsheet theory of the fundamental string as discussed in Section \ref{sec:near_BPS}. Importantly, the corresponding S-dual theory, which follows from a BPS decoupling limit of the D-string, is described by two-dimensional SYM.%
\footnote{In \cite{Blair:2023noj,Gomis:2023eav,Blair:2024aqz} this decoupling limit is given the name M1T (and more generally M$p$T for D$p$-branes) as it is implemented on arbitrary curved backgrounds in the full type II string theory. For recent studies of the S-duality between M1T and NRST see Refs.~\cite{Blair:2023noj, Ebert:2021mfu, Ebert:2023hba} as well as Refs.~\cite{Bergshoeff:2022iss,Bergshoeff:2023ogz} in which the $SL(2\,,\mathbb{Z})$ self-duality of NRST is investigated.}  
 In flat space, a stack of coinciding D-strings  is then described by the two-dimensional SYM action~\cite{Dijkgraaf:1997vv}
\begin{equation} \label{eq:matrixstring} 
S^{}_{\text{MST}}  = - \int \dd^2 \sigma \, \tr \Bigl(
  \tfrac{1}{2} \, \p_\alpha X^{i} \, \p^\alpha X^{i}
  + \tfrac{1}{4} \, g^{-2}_\text{YM} \, F_{\alpha \beta} \, F^{\alpha \beta} 
   - \tfrac{1}{4} \, g^{2}_\text{YM} \, \bigl[ X^{i}, X^{j} \bigl]^2
 \Bigr) \ . 
\end{equation} 
Here the YM coupling is related to the NRST coupling via $g^{}_\text{YM} = 1/(\sqrt{\alpha'} g_\text{s})$, so that in the IR (at strong YM coupling) the theory flows to free non-relativistic strings. 
Considering the matrix string action~\eqref{eq:matrixstring} in that limit restricts $X^i$ to be commuting, i.e. in the Cartan subalgebra of the $U(N)$ gauge group. We thus obtain a free superconformal field theory whose bosonic sector is described by  copies of non-relativistic strings, as we now review.%
\footnote{See  \cite{Harvey:1995tg} for possible subtleties in this argument.} 

Recall that matrix string theory is related to BFSS matrix theory via T-duality on a circle transverse to the D0-branes, the circle being T-dual to the compact direction $R_v$ on the worldsheet of NRST (after S-duality).  Since only the set of eigenvalues of $X^i$,  
 $x^i_I$, $I=1 \ldots N$, are gauge invariant, it follows that these may interchange when traversing the spatial circle of the worldsheet.  As a consequence, the fields $x^I (\sigma)$ can be multivalued, $x^i (\sigma + 2 \pi )= g x^i(\sigma) g^{-1} $
 with $g$ taking values in the Weyl group of $U(N)$ which is  the symmetric group $S_N$. We therefore get 
 an $S_N$ orbifold field theory, whose Hilbert space is decomposed into twisted sectors labeled by the conjugacy classes of the orbifold group $S_N$~\cite{Harvey:1995tg, Bershadsky:1995vm}. 
  The length of the permutation cycle in a twisted sector corresponds to the winding number of a non-relativistic string. 
 It follows that the $S_N$ orbifold CFT at $g^{}_\text{s} = 0$ describes the multi-string states that form a non-interacting super-selection sector. 
 
This confirms the heuristic picture presented earlier, namely when taking into account the different partitions $\sum_n n N_n = N$, type IIB NRST is a really the theory of matrix strings.
We note that, except for the recent Ref.~\cite{Blair:2024aqz}, the developments in NRST have focussed on the case of a single winding mode $w=N$, i.e a single twisted sector of fixed length $w$. 

In accordance with the general picture described earlier, interactions  can then be turned in matrix string theory by introducing twist field operators that join or split the strings at the positions where two eigenvalues of $X^i$ coincide. Therefore, matrix string theory at  finite $g^{}_\text{s}$ describes a second quantisation of NRST. As shown in \cite{Dijkgraaf:1997vv}, the coupling to the twist interaction term involves a coupling constant $\lambda \sim g_s \sqrt{\alpha'}$. By taking into account the combinatorial factor of the interaction term, we have that the NRST  worldsheet theory is perturbative for worldsheet energies satisfying
\begin{equation}
\label{pert_twist_int}
    E \ll \frac{1}{g_s \sqrt{w \alpha'}} \ . 
\end{equation}
Furthermore, in this second quantized formulation, the
string coupling becomes the worldsheet theory coupling. As a consequence, the topological worldsheet expansion arises due to the matrix-valued fields that interact via the twist operator, allowing for the joining and splitting of strings and hence topology change. 

Summarizing the above: Since all wound strings in the second quantized theory are wound along the same direction $v$, and the topological expansion can be thought of in terms of partitions of the winding along $v$, one can think of the second quantized theory as a single worldsheet theory. For $g_s=0$ this corresponds to partitions of the total winding number $w$ into wound strings, such that the total winding adds up to $w$. Each wound string is described by \eqref{NRST_worldsheet}. This is also known as $S_w$ orbifold CFT. Turning on $g_s$, these partitions can interact by splitting or joining, with total winding number conserved. Thus, this interacting theory will be what we refer to as {\sl NRST worldsheet theory} below.

\section{Phases of non-relativistic string worldsheet theory}
\label{sec:strong_coupling}

In this section we study the non-relativistic string theory (NRST) limit of the fundamental string soliton, being the backreacted winding mode of Section \ref{sec:near_BPS}. In Section \ref{sec:limit_Fstring_soliton} we show 
the backreaction blows up and that this means the limit is equivalent to a near-horizon limit. 
In Section \ref{sec:validity_Fstring_soliton} we show the near-horizon fundamental string soliton can be interpreted as a strong coupling phase of the NRST worldsheet theory (see Section \ref{sec:twisted_sectors}). In Section \ref{sec:phases_NRST_worldsheet} we furthermore explore the phases of the worldsheet theory at even stronger coupling.

\subsection{Limit of the fundamental string soliton}
\label{sec:limit_Fstring_soliton}

The gravitational soliton for the fundamental string (F-string) is a solution of equations of motion (EOMs) from the action that combines both the gravitational background fields and the F-string action which sources these fields
\begin{equation}
\label{fullI}
    I = I_{\rm gravity} + I_{\rm string} \ . 
\end{equation}
For the gravitational action $I_{\rm gravity}$, the relevant part is the NS-NS sector 
\begin{equation}
    I_{\rm gravity}  = \frac{1}{16\pi G} \int d^{10} x  \sqrt{-g} e^{-2\Phi} \left( R + 4 \partial_\mu \Phi \partial^\mu \Phi - \frac{1}{12} (dB)^2 \right) + \cdots  \ , 
\end{equation}
where $R$ is the curvature scalar of the ten-dimensional metric. The action $I_{\rm string}$ is the Polyakov action including Kalb-Ramond and dilaton term coupling to the background. The gravitational coupling and F-string tension are
\begin{equation}
    G = 8 \pi^6 (\alpha')^4 g_s^2 \spa T_{\rm F1} = \frac{1}{2\pi \alpha'} \ . 
\end{equation}

We assume an F-string with winding number $w>0$ along the spatial direction $v$.  This configuration is spherically symmetric in the transverse directions, and with Poincar\' e symmetry along the two longitudinal directions. With this as input, one can solve the EOMs of $I_{\rm gravity}$ away from the source, with the result \cite{Dabholkar:1995nc}
\begin{equation}
\label{Fstring_sol1}
ds^2 = c^2 H^{-1} (- dt^2 + dv^2 ) + dr^2 + r^2 d\Omega_7^{\, 2}
\spa B = - c^2 H^{-1} dt \wedge dv \spa g_s e^{\Phi} = g_s c \, H^{-1/2} \ , 
\end{equation}
where $H(r)$ is a harmonic function for the transverse directions $r^i$. We have imposed the boundary condition that this solution should asymptote to the flat background \eqref{solution_NRST}, as this corresponds to the NRST limit that we wish to take. This requires the boundary condition $H\rightarrow 1$ for $r\rightarrow \infty$. One can now include the source term in the EOMs from the full action \eqref{fullI} to find \cite{Dabholkar:1995nc}
\begin{equation}
\label{Fstring_sol2}
    H = 1+ c^2 \frac{L^6}{r^6} \ , 
\end{equation}
with
\begin{equation}
\label{Fstring_sol3}
L^6 = 32\pi^2 w\, g_s^2 \alpha'{}^3 \ . 
\end{equation}
The $c^2 L^6$ arise from the relative coupling 
\begin{equation}
\label{gravcoupl_Fstring}
   c^2   G\, T_{\rm F1}  \ , 
\end{equation}
between the source $I_{\rm string}$ and the background $I_{\rm gravity}$. Here the $c^2$ factor appears from the asymptotics of $e^{2\Phi}$. Note also the winding number is equated with the amount of charge of the F-string \cite{Dabholkar:1995nc}. 

In the $c\rightarrow \infty$ limit the $c$ factor of the harmonic function \eqref{Fstring_sol2} combines with the $c$ factors in \eqref{Fstring_sol1} to the following finite background 
\begin{equation}
\label{Fstring_NH}
ds^2 = \frac{r^6}{L^6} (- dt^2 + dv^2 ) + dr^2 + r^2 d\Omega_7^{\, 2}
\spa B = -  \frac{r^6}{L^6}  dt \wedge dv \spa g_s e^{\Phi} = g_s  \frac{r^3}{L^3}  \ . 
\end{equation}
As first noticed in \cite{Avila:2023aey}, this is the near-horizon limit of the F-string soliton. Thus, the NRST limit becomes equivalent to a near-horizon limit for this particular gravitational soliton of string theory. 
This is due to the fact that the gravitational coupling to the F-string \eqref{gravcoupl_Fstring} goes like $c^2$ and thus blows up in the $c\rightarrow \infty$ limit. Hence, the asymptotic region where $H\simeq 1$ is abolished, due to the infinite strength of the backreaction of the F-string.

\subsection{Regime of validity of the near-horizon fundamental string soliton}
\label{sec:validity_Fstring_soliton}

We saw above that the F-string soliton in the $c\rightarrow \infty$ limit gives a background of relativistic string theory rather than NRST \cite{Avila:2023aey}. 
However, as we shall see below, the correct interpretation of this fact is that the near-horizon F-string soliton provides a description of the non-perturbative phase of the 1+1 dimensional NRST worldsheet theory (as defined in Section \ref{sec:twisted_sectors}).
The near-horizon F-string soliton \eqref{Fstring_NH} is a valid semi-classical description provided the curvature length scale of the background is much bigger than the string length 
\begin{equation}
\label{small_curv_Fstring}
    r\gg \sqrt{\alpha'} \ , 
\end{equation}
and the string coupling in \eqref{Fstring_NH} is weak 
\begin{equation}
g_s \frac{r^3}{L^3} \ll 1 \ . 
\end{equation}
Using \eqref{Fstring_sol3} this implies \cite{Itzhaki:1998dd}
\begin{equation}
    1 \ll \frac{r}{\sqrt{\alpha'}} \ll w^{\frac{1}{6}} \ , 
\end{equation}
which means in particular that the winding number needs to be large $w \gg 1$. 

We can interpret \eqref{small_curv_Fstring} in terms of the energy scale of the worldsheet theory. Indeed, following \cite{Peet:1998wn} one considers the s-wave mode of a massless scalar field on the background \eqref{Fstring_NH}. Write
\begin{equation}
    \psi( t,v,r ) = e^{-iEt} F(r) \ , 
\end{equation}
where $E$ is the worldsheet energy and we have set the worldsheet momentum to zero.
Then we have
\begin{equation}
    \frac{L^6}{r^4} E^2 F + r^{-11} \partial_r ( r^{13} \partial_r F) =0 \ . 
\end{equation}
This shows the typical radial scale is $r \sim \sqrt{L^3 E}$. Thus, for a given radius $r$ the typical worldsheet energy scale is
\begin{equation}
    E \sim \frac{r^2}{L^3} \sim \frac{r^2}{\sqrt{w}g_s (\alpha')^{\frac{3}{2}}} \ . 
\end{equation}
Therefore, the small curvature condition \eqref{small_curv_Fstring} is equivalent to
\begin{equation}
\label{small_curv_Fstring_energy}
    E \gg \frac{1}{\sqrt{w}g_s\sqrt{\alpha'}} \ . 
\end{equation}
In case NRST arises from a $c\rightarrow \infty$ limit of type IIB relativistic string theory, this is in accordance with the worldsheet theory of NRST being perturbative for \eqref{pert_twist_int}. 
Hence, we can interpret the onset of the weak curvature condition \eqref{small_curv_Fstring_energy} as the point for which the NRST worldsheet theory becomes strongly coupled. In other words, the backreaction of the F-string soliton in the NRST limit only sets in at this point.
Thus, strong coupling and high energy go hand-in-hand for the NRST worldsheet theory.
In conclusion, this means that when the NRST worldsheet theory is weakly coupled, it describes strings coupling to a TSNC geometry. 

Finally, we note that in the weak coupling regime $w g_s^2 \ll 1$ there is a window of energies above the string scale $1/\sqrt{\alpha’} \ll E \ll 1/(\sqrt{w}g_s \sqrt{\alpha’})$. Thus, it is possible to consider the tower of massive string states in this regime.


\subsection{Strong coupling phases of NRST worldsheet theory}
\label{sec:phases_NRST_worldsheet}

Beyond $r \gg w^{\frac{1}{6}} \sqrt{\alpha'}$ the strong-coupling/high-energy phases are depending on whether one approaches NRST from the $c\rightarrow \infty$ limit of type IIA or IIB relativistic string theory.

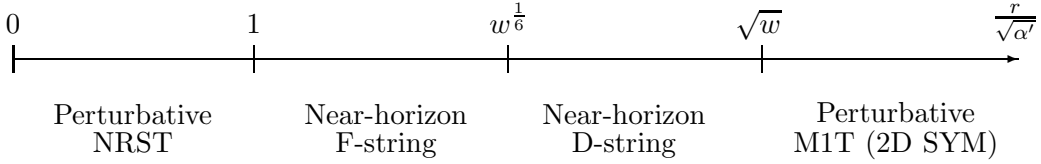
\begin{figure}[h]
\begin{picture}(390,75)(-25,0)
\put(5,40){\vector(1,0){380}}
\put(375,53){$\frac{r}{\sqrt{\alpha'}}$}
\put(20,5){\shortstack{Perturbative \\ NRST}}
\put(115,5){\shortstack{Near-horizon \\ F-string}}
\put(205,5){\shortstack{Near-horizon \\ D-string}}
\put(300,5){\shortstack{Perturbative \\ M1T (2D SYM)}}
\put(2,50){0}
\put(5,35){\line(0,1){10}}
\put(93,50){$1$}
\put(96,35){\line(0,1){10}}
\put(185,50){$w^{\frac{1}{6}}$}
\put(192,35){\line(0,1){10}}
\put(278,50){$\sqrt{w}$}
\put(288,35){\line(0,1){10}}
\end{picture}
\caption{Phase diagram for the NRST limit of type IIB string theory. \label{fig:IIB_phases} }
\end{figure}

Starting with type IIB relativistic string theory, the first phase for $r \gg w^{\frac{1}{6}} \sqrt{\alpha'}$ one enters is the near-horizon D-string soliton. This is obtained by S-dualizing the near-horizon F-string soliton \eqref{Fstring_NH}, giving the background
\begin{equation}
ds^2 = g_s^{-1} \frac{r^3}{L^3} \left[ - dt^2 + dv^2  + \frac{L^6}{r^6} ( dr^2 + r^2 d\Omega_7^{\, 2} ) \right]
\spa 
A_{(2)} = - \frac{r^6}{L^6} dt \wedge dv \spa \frac{1}{g_s} e^{\Phi} = \frac{L^3}{g_s r^3}   \ . 
\end{equation}
This is easily seen to be valid for \cite{Itzhaki:1998dd}
\begin{equation}
    w^{\frac{1}{6}} \ll \frac{r}{\sqrt{\alpha'}} \ll \sqrt{w} \ , 
\end{equation}
where the lower bound on $r/\sqrt{\alpha'}$ corresponds to weak string coupling while the upper bound is weak curvature. 
Finally for $r \gg \sqrt{w}\sqrt{\alpha'}$, one enters the phase of weakly coupled super Yang-Mills (SYM) with action \eqref{eq:matrixstring} \cite{Motl:1997th,Dijkgraaf:1997vv,Itzhaki:1998dd}, also known as M1T \cite{Blair:2023noj,Gomis:2023eav}. 
We have depicted the phase diagram of NRST arising from the $c\rightarrow \infty$ limit type IIB in Figure \ref{fig:IIB_phases}.

Turning to type IIA relativistic string theory, the phase for $r \gg w^{\frac{1}{6}} \sqrt{\alpha'}$ one enters is the near-horizon M2-brane soliton, given by 
\begin{equation}
\label{Fstring_NHA}
ds^2 = \frac{r^4}{L^4} (- dt^2 + dv^2 + dx_{11}^{\, 2} ) + \frac{L^2}{r^2} dr^2 + L^2 d\Omega_7^{\, 2}
\spa C_{(3)} = -  \frac{r^6}{L^6}  dt \wedge dv \wedge dx_{11}  \ , 
\end{equation}
with Planck length given by $l_p^3 = g_s (\alpha')^{\frac{3}{2}}$ and period of $x_{11}$ being $2\pi R_{11} = 2\pi g_s \sqrt{\alpha'}$.
This background has weak curvature for $L \gg l_p$ corresponding to $w \gg 1$. Moreover, to be in M-theory, the effective radius of $x_{11}$ should be much larger than the Planck length $\tfrac{r^2}{L^2} R_{11} \gg l_p$ which corresponds to $r \gg w^{\frac{1}{6}} \sqrt{\alpha'}$. Note that the near-horizon M2-brane soliton is holographically dual to the three-dimensional maximally superconformal field theory in three dimensions.
We have depicted the phase diagram of NRST arising from the $c\rightarrow \infty$ limit type IIA in Figure \ref{fig:IIB_phases}.

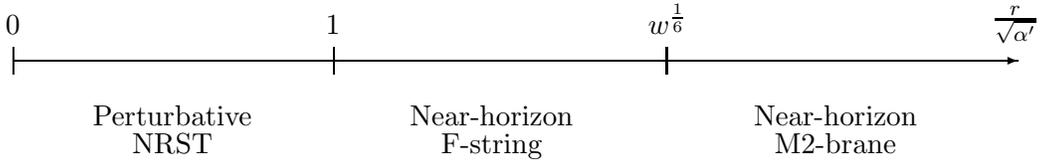
\begin{figure}[h]
\begin{picture}(390,75)(-25,0)
\put(5,40){\vector(1,0){380}}
\put(375,53){$\frac{r}{\sqrt{\alpha'}}$}
\put(35,5){\shortstack{Perturbative \\ NRST}}
\put(155,5){\shortstack{Near-horizon \\ F-string}}
\put(285,5){\shortstack{Near-horizon \\ M2-brane}}
\put(2,50){0}
\put(5,35){\line(0,1){10}}
\put(123,50){$1$}
\put(126,35){\line(0,1){10}}
\put(245,50){$w^{\frac{1}{6}}$}
\put(252,35){\line(0,1){10}}
\end{picture}
\caption{Phase diagram for the NRST limit of type IIA string theory. \label{fig:IIA_phases} }
\end{figure}

We note that the above analysis of the various phases encountered in the NRST limit is reminiscent of the analysis of the D-brane decoupling limits \cite{Itzhaki:1998dd} as well as  the BFSS matrix model limit \cite{Polchinski:1999br}.

\section{Solitons and the non-relativistic string limit}
\label{sec:branes_NRST}

In this section we turn to gravitational solitons of NRST in the regime where it is described by a bulk gravitational theory, albeit a type of non-Lorentzian gravity theory.%
\footnote{See \cite{Hartong:2022lsy} for a review on Newton-Cartan geometry and gravity for particles.} 
In fact, following the construction of the worldsheet action of non-relativistic strings in string Newton-Cartan backgrounds \cite{Harmark:2017rpg,Bergshoeff:2018yvt,Bidussi:2021ujm}, 
corresponding low-energy effective actions for this sector have been obtained in \cite{Gomis:2019zyu, Gallegos:2019icg, Yan:2019xsf, Bergshoeff:2021bmc, Yan:2021lbe} using various methods. The resulting  actions take the form of dynamical string Newton-Cartan gravity theories. In practice these actions have so far only been computed partially, e.g. the RR sector in its most general form is not known from a first principle computation. However, conceptually it should be clear that one expects such actions to have interesting gravitational solitons, in parallel with the existence of D$p$ and NS5-branes in relativistic string theory. 

Since, as explained in Section \ref{sec:near_BPS},  NRST can be obtained as a geometric limit from relativistic string theory, one expects that one way to generate such gravitational solitons is by implementing this limit on appropriate solutions of the relativistic low-energy
effective action.  While we have just seen that this does not work for the F-string, we will show in this section that there
are gravitational solitons that do survive the NRST limit. As a specific example, we will first illustrate this for transverse D-branes , employing the
DLCQ/T-duality procedure of Section \ref{sec:near_BPS}. We then give a more general procedure for generating such NRST solutions and illustrate this by applying the procedure to a second example, the NS5-brane. 
As a consequence we find that transverse D-branes and NS5-branes survive.%
\footnote{Analogous examples of non-Lorentzian background solutions in matrix $p$-brane theory, which comprise  U-dual corners of NRST, were constructed in \cite{Lambert:2024yjk,Blair:2023noj,Lambert:2024ncn}.} 
We end the section by discussing the near-horizon geometries of such NRST solutions, focussing in particular on the transverse D-branes, and their role in novel holographic dualities.

\subsection{Non-relativistic D-brane solitons \label{sec:NRD} }

To generate D-brane solutions in NRST, we start with the well-known solution for a D$(p+1)$-brane
in type II supersymmetric string theory. 
\begin{equation}
\label{Dp+1brane}
ds^2 = H^{-\frac{1}{2}} \left(- dT^2 + dX^2 + \sum_{a=1}^{p} (dy^a)^2 \right)+ H^{\frac{1}{2}} \sum_{i=1}^{8-p} dr^i dr^i \ , 
\end{equation}
\begin{equation}
A_{(p+2)} = (H^{-1} - 1) dT \wedge dX \wedge dy^1 \wedge \cdots  \wedge dy^{p}
\spa 
\bar{g}_s e^{\Phi} = \bar{g}_s H^{\frac{2-p}{4}} \ , 
\end{equation}
\begin{equation}
\label{HDbrane} 
H = 1 + \frac{(2\pi \sqrt{\alpha'})^{6-p} \bar{g}_s N}{(6-p) \Omega_{7-p} r^{6-p}} \ . 
\end{equation}
Here we have singled out the coordinates $(T,X)$ on the worldvolume to which we apply the coordinate transformation
\eqref{coords_tu}, yielding 
\begin{equation}
ds^2 = H^{-\frac{1}{2}} \left( 2dt\, du + \frac{1}{c^2}{du^2} + \sum_{a=1}^{p} (dy^a)^2 \right)+ H^{\frac{1}{2}} \sum_{i=1}^{8-p} dr^i dr^i \ , 
\end{equation}
\begin{equation}
A_{(p+2)} = (H^{-1} - 1) dt \wedge du \wedge dy^1 \wedge \cdots  \wedge dy^{p} 
\spa 
\bar{g}_s e^{\Phi} = \bar{g}_s H^{\frac{2-p}{4}} \ . 
\end{equation}
We subsequently do a T-duality in the $u$-direction, giving%
\footnote{Note that this solution is smeared along the $v$ isometry. See \cite{Guijosa:2023qym,Avila:2023aey} for a study of the NRST limit in the non-smeared case.} 
\begin{equation}
\label{DbraneTa}
ds^2 = H^{-\frac{1}{2}} \left( -c^2 dt^2 + \sum_{a=1}^{p} (dy^a)^2 \right)+ H^{\frac{1}{2}} \left( c^2 dv^2 +  \sum_{i=1}^{8-p} dr^i dr^i \right)  \ , 
\end{equation}
\begin{equation}
\label{DbraneTb}
B = c^2 dt \wedge dv
\spa
A_{(p+1)} = (H^{-1} - 1) dt \wedge dy^1 \wedge \cdots  \wedge dy^{p}
\spa 
g_s e^{\Phi} = g_s c \, H^{\frac{3-p}{4}} \ . 
\end{equation}
We note that in the expression above the dilaton is the transformed field  $\Phi_{v}$, but for brevity we omit the subscript. Moreover, the harmonic function takes the form
\begin{equation}
\label{HNRD}
    H = 1 + \frac{(2\pi \sqrt{\alpha'})^{7-p} g_s N}{(6-p) \Omega_{7-p} R_v r^{6-p}} \ , 
\end{equation}
where we used \eqref{Ru_Rv}, \eqref{bar_gs}. 

Finally, we can take the NRST limit $c \rightarrow 0$ on this solution. As, we now show, this will lead to a string Newton-Cartan background. To this end we recall how this geometry can be read off in general when taking the NRST limit of a curved type II supergravity background \cite{Harmark:2017rpg,Bergshoeff:2018yvt,Bergshoeff:2021tfn,Bidussi:2021ujm,Ebert:2021mfu} 
\footnote{Note that the finite part in the $B$ field, denoted by $b$ in \eqref{eq:nrsl}, corresponds to the Newton-Cartan two-form $m$ identified in \cite{Bidussi:2021ujm}.}
\begin{equation} \label{eq:nrsl}
G^{}_{\mu\nu} = c^2 \, \tau^{}_{\mu\nu} + E^{}_{\mu\nu} \spa B  =  c^2 \, \tau^0 \wedge \tau^1 + b \spa  \tau^A = \tau_\mu{}^A \, dx^\mu  \ , 
\end{equation}
\begin{equation} \label{eq:nrsl2}
A_{(q)} = c^2 \, \tau^0 \wedge \tau^1 \wedge a_{(q-2)} + a_{(q)}  \spa 
    e^{\Phi} = c \, e^{\varphi}  \ , 
\end{equation}
where $A=0,1$ runs over the tangent space of the longitudinal directions, 
$\tau_{\mu \nu} = \tau^A_\mu \tau^b_\nu \,\eta_{AB}$ and $E_{\mu \nu} = 
E_\mu^i E_\nu^j \,\delta_{ij} $ the transverse metric. 
The string-Newton Cartan geometry then is described by the background fields
$\tau_{\mu \nu} $, $E_{\mu \nu}$, $\varphi$ along with
the further fields $b$, $a_q$, $a_{q-2}$ (which can possibly vanish). Note that the flat SNC background is given in \eqref{eq:flatTSNC}. 
We also emphasize that this is a non-Lorentzian background since the  original $\mathrm{SO}(1,9)$ local Lorentz transformation is now broken into
 $SO(1,1)$ transformations acting on $\tau_\mu^A$, 
 $\mathrm{SO}(8)$ transformation acting on $E_\mu{}^i$ along with string Galilean boosts, parameterized by $\Lambda^i_A$ that
 transform  $\delta \tau_\mu{}^A = 0$ 
 and $\delta E_\mu{}^i = \Lambda_A^i \, \tau_\mu{}^A$. The field $b$ also transforms under these \cite{Bidussi:2021ujm}. 
 
Comparing  \eqref{DbraneTa}, \eqref{DbraneTb} with \eqref{eq:nrsl}, \eqref{eq:nrsl2}, then yields the following 
new TSNC background 
\begin{equation}
\label{NRDbrane} 
\tau_{\mu\nu} dx^\mu dx^\nu = -H^{-\frac{1}{2}}  dt^2 + H^{\frac{1}{2}} dv^2\spa 
E_{\mu\nu} dx^\mu dx^\nu  = H^{-\frac{1}{2}}  \sum_{a=1}^{p} (dy^a)^2 + H^{\frac{1}{2}}   \sum_{i=1}^{8-p} dr^i dr^i  \ ,
\end{equation}
\begin{equation}
a_{(p+1)} = (H^{-1} - 1) dt \wedge dy^1 \wedge \cdots  \wedge dy^{p}
\spa 
g_s   e^{\varphi} =  g_s H^{\frac{3-p}{4}} \ , 
\end{equation}
with the harmonic function given in \eqref{HNRD}.  
One can presumably view this as the backreacted version of a non-relativistic D$p$-brane. It would be interesting to make
this more precise using boundary state computations, in analogy with that of the relativistic case.

We note that in the limiting procedure the harmonic function  stayed finite in this case. One physical way to see this is 
via the fact that the tension of the D$p$ brane times the effective 9D gravitational coupling remains finite in the limit, since all factors of
$c$ cancel in the product, 
\begin{equation}
{\rm T}_{{\rm D}p} \cdot G_{\rm N} =  \frac{N}{(2\pi)^p (\alpha')^{\frac{p+1}{2}} g_s c} \cdot \frac{8 \pi^6 (\alpha')^4 g_s^2 c^2}{c R_v}  \sim c^0 \ . 
\end{equation}

We also emphasize that these D-branes are transverse to the  $v$-direction along which the F-string is wound (before the limit), which is why we call them transverse D-branes.  In fact, a related observation is that the relativistic F1-D$p$ bound state (with F1 transverse to the D$p$-brane) is a 1/4-BPS bound state in relativistic string theory. The mass of this is thus given by the sum  $M_{{ \rm F}_1}$ + $M_{{\rm D}p}$. Since the non-relativistic string limit involves a critical field cancelling the $M_{{ \rm F}_1}$ part, the resulting bound state gives a finite mass excitation after the limit \cite{bpslimits}.%
\footnote{A U-dual version of this observation, namely that D0-branes survive the M4T limit and 
D4-branes survive the M0T limit was made in \cite{Blair:2024aqz}.}

Finally, we remark that the TSNC D-brane geometry has non-trivial longitudinal vielbeine $\tau_\mu^A$. It would be interesting to
examine this in light of the results \cite{Bergshoeff:2021tfn,Bergshoeff:2024nin} in which consistency
of the non-relativistic limit with 10D/11D supersymmetry, leads to certain constraints on the bosonic intrinsic torsion tensors along with additional requirements  (see also \cite{Yan:2021lbe}). 

\subsection{General procedure for NRST solutions and application to NS5-brane \label{sec:proc}}

From the explicit construction of the transverse NR D-brane in the previous section, we can infer a general procedure to generate such NRST geometries, which can be used to capture further generalizations. 

We can  generate NRST solutions from  any background of relativistic string theory, that solves the type II supergravity equations of motion, as long as it  has the property that it  has $ISO(1,1)$ isometry. In addition, denoting the two translational Killing symmetries in $ISO(1,1)$ by $P_0^\mu$, $P_1^\mu$, we need to impose the requirement
$B_{ \mu \nu} P_0^\mu P_1^\mu =0$ for the Kalb-Ramond $B$ field.  The latter means that one can choose coordinates $(T,X,r^i)$ such that 
\begin{itemize}
    \item The fields of the background do not depend on $T$ and $X$.
    \item $-g_{TT}=g_{XX}$ and $g_{TX}=0$.
    \item $B_{TX}=0$.
\end{itemize}
As an illustration, this is clearly the case for the D-brane seed solution \eqref{Dp+1brane}.  Given a solution of this type, it is obvious  that one can perform the procedure applied in Section \ref{sec:near_BPS}, i.e. transforming to the $(t,u)$ system and subsequently T-dualizing along $u$ and taking the infinite $c$ limit.  We remark that any  solution generated this way
will have isometries along the $t$ and $v$ directions.

Furthermore, clearly the 
the longitudinal F-string soliton  does not obey the requirements above, which is why we cannot use it to generate a NRST solution. 
  Another remark, to which we will return in the outlook, is that this immediately shows that the procedure does not work for non-extremal solutions, since they have a blackening factor that destroys a possible rotational invariance between $T$ and $X$.

Returning to the solution generating technique, in the coordinates $(T,X,r^i)$ this means that the seed background should be of the form
\begin{equation}
ds^2 = f(r) [ -dT^2 + dX ^2]
+ g_{ij} (r) dr^i dr^j  \ , 
\end{equation}
\begin{equation}
B = B_{ij} (r) dr^i \wedge dr^j \spa e^{\Phi} = g(r) \ , 
\end{equation}
where we emphasize that here $r $ denotes the set of transverse coordinates $r^i$.  In addition, we need that any non-zero RR background fields should include either both the longitudinal directions or none of them, while they should also be dependent on only $r^i$. 

Using the procedure of Section \ref{sec:near_BPS} and the identification \eqref{eq:nrsl} for the TSNC geometry, the relativistic background generates the following NRST background
\begin{equation}
\label{NRST_solution} 
\tau_{\mu\nu} dx^\mu dx^\nu = -f(r) dt^2 + f(r)^{-1}  dv^2\spa 
E_{\mu\nu} dx^\mu dx^\nu  = 
g_{ij } (r)  dr^i dr^j  \ , 
\end{equation}
\begin{equation}
b_{(2)} = B_{ij} (r) dr^i \wedge dr^j \spa 
e^{\varphi} = g(r) \ ,
\end{equation}
along with  any RR potentials that are present in the seed solution. In addition, any $\bar{g}_s$ dependence in the seed solution becomes $ g_s \sqrt{\alpha'}/R_v$ in the NRST solution. 

As a concrete application, one can take the relativistic NS5-brane solution solution as a seed solution 
\begin{equation}
    ds^2 = -dT^2 +dX^2 + \sum_{m=1}^4 dy_m^2 + H \sum_{i=1}^4 dr_i^2 \spa
        (dB)_{ijk} = \varepsilon_{ijk}{}^l H\partial_l H \spa  \bar g_s e^\Phi = \bar g_s H^{1/2} \ , 
\end{equation}
\begin{equation}
H = 1 + \frac{N\alpha'}{r^2} \ ,
\end{equation}
with $r^2 = \sum_{i=1}^4 (r^i)^2$.
We note that this involves the $(T,X)$ coordinates on the worldvolume of the NS5-brane.  We can immediately read off from \eqref{NRST_solution} the resulting TSNC geometry 
\begin{equation}
\tau_{\mu\nu} dx^\mu dx^\nu = - dt^2 + dv^2\spa 
E_{\mu\nu} dx^\mu dx^\nu =  \sum_{m=1}^4 dy_m^2 + H \sum_{i=1}^4 dr_i^2 \ ,
\end{equation}
\begin{equation}
        (db)_{ijk} = \varepsilon_{ijk}{}^l H\partial_l H \spa  g_s e^{\varphi} = g_s H^{1/2}
        \spa 
        H = 1 + \frac{N\alpha'}{r^2} \ .
\end{equation}
This solution coincides with the one given in Ref.~\cite{Bergshoeff:2022pzk}, where it was found using other methods, and shown to be $1/2$-BPS. Here we thus see how it relates to the original relativistic NS5-brane solution. It is interesting to note that, contrary to the NRST D-brane solution,  due to the flatness on the longitudinal part of the metric this solution has zero torsion.

In analogy with the D-brane case discussed above, the finiteness of the harmonic function in the limit
is again a consequence of the fact that 
the tension of the NS5 brane times the effective 9D gravitational coupling remains finite
\begin{equation}
{\rm T}_{{\rm NS}5} \cdot G_{\rm N} 
= \frac{N}{(2\pi)^5 (\alpha')^3 g_s^2 c^2} \cdot 8 \pi^6 (\alpha')^4 g_s^2 c^2  \sim c^0 \ .
\end{equation}
Furthermore, we see that this transverse NR NS5-brane background is obtained from the 1/4-BPS NS5-F1 bound state.

\subsection{Near-horizon limit and holographic duality with bulk NRST}

Now that we have established the existence of various solitons in NRST, 
a natural question is whether one can generate novel holographic dualities
involving NRST in the bulk by taking a further near-horizon limit.
This was recently shown in the context of matrix theory \cite{Blair:2024aqz}, leading to the insight that holographic dualities in string theory can be viewed as DLCQ${}^m$/DLCQ${}^n$ correspondences with $m<n$
where the extra DLCQ on the boundary corresponds to the near-horizon limit in the bulk. 

For definiteness we focus here on the NRST D-brane solution. 
The question is thus how to  implement a further decoupling limit such that it acts as a near-horizon limit on
the string Newton-Cartan background \eqref{NRDbrane} describing the NR D-brane solution.  Following the paradigm of \cite{Blair:2024aqz}, we expect that this can be achieved by doing a ``2nd DLCQ'' in the asymptotic region. 
More precisely, it was shown in \cite{Blair:2024aqz} that the near-horizon limit of a bulk D$q$-brane solution corresponds to the M$q$T   limit in the asymptotic region. The latter involves the BPS decoupling limit of a stack of D$q$-branes, and is in fact 
a U-dual analogue of the NRST limit, all of which lie in the same DLCQ-orbit. 

For the case at hand, since we started with a D$(p+1)$-brane solution in \eqref{Dp+1brane}, this means that if we first
perform the M$(p+1)$T limit, it will implement the near-horizon limit, i.e. yielding the same form of the solution \eqref{Dp+1brane} but now with the near-horizon form, i.e. the ``1'' omitted in the harmonic function. 
We can then subsequently repeat the same steps as in Section \ref{sec:NRD} to do the NRST limit, ending up with the near-horizon limit of \eqref{NRDbrane}.
This SNC solution takes the form 
\begin{equation}
\label{nhNRDbrane} 
\tau_{\mu\nu} dx^\mu dx^\nu = - 
\left( \frac{r}{\ell} \right)^{(6-p)/2 }  dt^2 + \left( \frac{r}{\ell} \right)^{-(6-p)/2 } dv^2 \ , 
\end{equation}
\begin{equation} 
E_{\mu\nu} dx^\mu dx^\nu  =  \left( \frac{r}{\ell} \right)^{(6-p)/2 } 
\sum_{a=1}^{p} (dy^a)^2 + \left( \frac{r}{\ell} \right)^{-(6-p)/2 }    \sum_{i=1}^{8-p} dr^i dr^i  \ ,
\end{equation}
\begin{equation}
a_{(p+1)} =  \left( \frac{r}{\ell} \right)^{6-p }  dt \wedge dy^1 \wedge \cdots  \wedge dy^{p}
\spa 
  g_s e^{\varphi} =   g_s \left( \frac{r}{\ell} \right)^{-(6-p)(3-p)/4 }  \ ,
\end{equation}
where we have defined 
\begin{equation}
\ell^{6-p} =  \frac{(2\pi \sqrt{\alpha'})^{7-p} g_s N}{(6-p) \Omega_{7-p} R_v } \ .
 \end{equation}
 Interestingly, we know  how to describe the resulting holographic dual theory
on the boundary is. It corresponds to  M$(p+1)$T followed by an additional DLCQ/T-duality (NRST limit). 
But this theory is precisely what was dubbed MM$p$T in \cite{Blair:2023noj,Gomis:2023eav}. In particular the
worldvolume formulation of this theory was studied in \cite{Gomis:2023eav}, to which we refer for the actions of the fundamental string and D$p$-brane respectively. The curved non-Lorentzian target space (called MM$p$T geometry) was obtained and studied in \cite{Blair:2023noj,Gomis:2023eav}. Importantly, we can view this theory as arising from  a multicritical BPS decoupling limit involving the F1-D$p$ brane bound state.      It would be interesting to
 study this MM$p$T boundary theory in more detail, which we leave for future work.   
 
 We also note that, the near-horizon limit of the NRST NS5-brane solution 
 \eqref{nhNRDbrane} can be similarly obtained by doing a ``2nd DLCQ''. In particular, we expect it to correspond to a dual boundary theory obtained from the multicritical F1-NS5 brane decoupling limit.  It would be interesting to study this further.

\section{Discussion}
\label{sec:discussion}

In this paper we explore the role of gravitational solitons in non-relativistic string theory (NRST). 
A main conclusion, is that far from all solitons of type II string theory will become solitons of NRST, upon taking the NRST limit. 

Indeed, considering the fundamental string soliton, corresponding to a backreacted winding mode, the NRST limit becomes equivalent to the near-horizon limit of that soliton, signifying that the gravitational backreaction of the soliton blows up in the limit \cite{Avila:2023aey}. 
We show that this can be interpreted as describing a strong coupling phase of the worldsheet theory of NRST. This is supported by the connection between NRST and matrix string theory \cite{Motl:1997th,Dijkgraaf:1997vv,Itzhaki:1998dd}, suggesting that when including multi-string states, the interactions between the multi-string states become strongly coupled precisely as the near-horizon phase takes over.

We find two classes of gravitational solitons of type II string theory that do survive the NRST limit to become gravitational solitons of NRST. To show this, we employ the same procedure of a near-lightcone transformation and a T-duality that we have seen explicitly can connect the NRST limit to a DLCQ/null-reduction limit of the string. This procedure is turned into a solution generator. 
For the NS5-brane soliton, the solution generator is applicable, and reveals a gravitational soliton of NRST in the NS-NS sector equivalent to a previously discovered supersymmetric soliton of NRST \cite{Bergshoeff:2022pzk}. 
Starting with a D$(p+1)$-brane, longitudinal to the string, and using the solution generator, one obtains novel gravitational solitons corresponding to transverse D$p$-branes in NRST. We interpret these as backreacted versions of the transverse D$p$-brane configurations of the open string sector of NRST \cite{Gomis:2020fui,Gomis:2020izd,Kluson:2020kyp,Ebert:2021mfu,
Guijosa:2023qym,Hartong:2024ydv}. 

Finally, we consider a second decoupling limit of the transverse D$p$-branes in NRST. As the NRST limit by itself can be seen as a DLCQ, this second decoupling limit thus corresponds to a second DLCQ of type II string theory. We show the second DLCQ limit is realized in the bulk as a near-horizon limit of the transverse D$p$-branes gravitational soliton. Moreover, we show that from a worldvolume point of view the second DLCQ precisely corresponds to the multicritical matrix limit (MM$p$T) of \cite{Harmark:2017rpg,Harmark:2018cdl,Blair:2023noj,Gomis:2023eav}. This provides a novel holographic correspondence between NRST on a TSNC near-horizon background and multicritical matrix theory. 

There are a number of interesting future directions to pursue. 
\begin{itemize}
\item It is of obvious interest to study and elucidate the proposed holographic correspondence between multicritical matrix theory and the near-horizon transverse D$p$-brane backgrounds of NRST \eqref{nhNRDbrane}. 
In particular, for the case $p=3$, it seems feasible to develop a  concrete holographic correspondence, on par with the AdS/CFT correspondence, for which the gravity side is based on TSNC geometry. 
\item It would be interesting to examine in more detail how the 
transverse D$p$-brane gravitational soliton \eqref{NRDbrane}
fits with torsional constraints arising from supersymmetry. Such constraints have been found in the formulation of a supergravity version of NRST and its M-theory equivalent \cite{Bergshoeff:2021tfn,Bergshoeff:2024nin}. For the case at hand, it could be useful to explore this via the finite-$c$ map  of Section \ref{sec:near_BPS}, which has also been used previously \cite{Bergshoeff:2019pij} to study the $c\rightarrow \infty$ of beta-functions of type II string theory. 
\item In type II string theory, one can promote the various gravitational solitons of BPS states to non-extremal backgrounds with non-zero temperatures. Our solution generating technique of Section \ref{sec:proc} suggests that such non-extremal backgrounds are not realized in NRST. It would be interesting to understand if this is tied to the non-Lorentzian nature of TSNC geometry, and also what the correct procedure is for studying non-zero temperature in NRST. 
\item
Following also \cite{Blair:2024aqz}, we have further elaborated on how including multi-string states in NRST reveals a clear connection to matrix string theory \cite{Motl:1997th,Dijkgraaf:1997vv}. 
This connection shows 
that it would be natural to expect the existence of a further generalization of the NRST sigma-model on a curved TSNC background \cite{Harmark:2017rpg,Bergshoeff:2018yvt,Harmark:2018cdl,Harmark:2019upf,Oling:2022fft} to a new matrix-valued sigma-model. This new matrix-valued sigma-model should be able to capture the multi-string states in NRST. In a flat background and with zero string coupling, it should reduce to free NRST strings, generically with multiple winding modes describing a multi-string state. Turning on the string coupling and for a curved background, the matrix-valued sigma-model should capture the splitting and joining interaction of multi-string states, including the interactions that arise from curvature of the background. This would in turn promote the background geometry to a matrix-valued TSNC geometry.
This suggests that one should view the matrix-valued version of TSNC geometry as a quantum geometry. If true, this would be a successful example of approaching the question of quantum geometry by considering the corner of the Bronstein cube with Planck's constant and Newton's constant finite and the speed of light sent to infinity. 
\item Finally, we highlight that it was shown in \cite{Blair:2024aqz} that when viewed in reverse, BPS decoupling limits can be related to the $T \bar{T}$ deformation. This general insight built on the observation \cite{Blair:2020ops} that the particular BPS decoupling limit defining NRST  can be viewed in reverse as the $T \bar T$ deformation of a two-dimensional field theory. While this is well-understood for a single string,  it would be very interesting to try to generalize this statement for the matrix valued fields appearing in the second quantized from of NRST. 
More generally, there could be connections in this respect to the large speed of light expansions (as opposed to limit)
of string theory, pursued in \cite{Hartong:2021ekg,Hartong:2022dsx} following the analogous expansions of GR considered in 
\cite{Hansen:2019pkl,Hansen:2020pqs}. 

\end{itemize}

\section*{Acknowledgments}

We would like to thank Chris Blair,  Jan de Boer, Emil Have, Jelle Hartong and Ziqi Yan for useful discussions.  
T.H. thanks Nordita for the support as corresponding fellow. 
The work of N.O. is supported in part by VR project Grant 2021-04013 and Villum Foundation Experiment Project No.~00050317.  Nordita is supported in part by NordForsk.

\addcontentsline{toc}{section}{References}

\bibliography{newbib,mtr}
\bibliographystyle{newutphys}

\end{document}